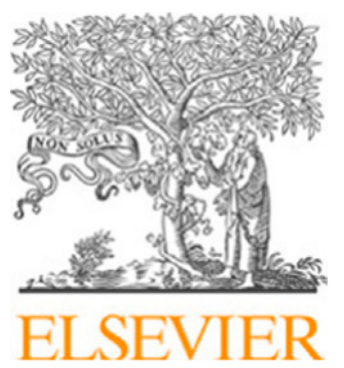

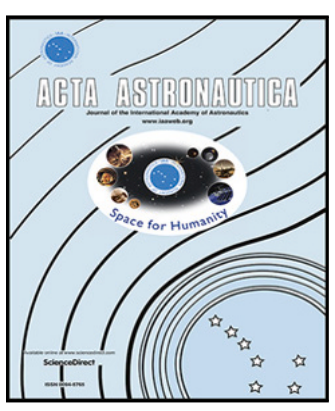

Research paper

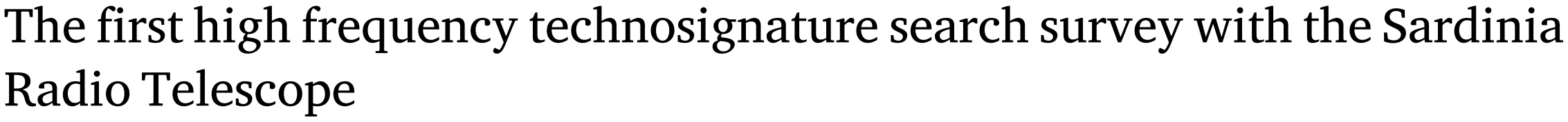

# The first high frequency technosignature search survey with the Sardinia Radio Telescope

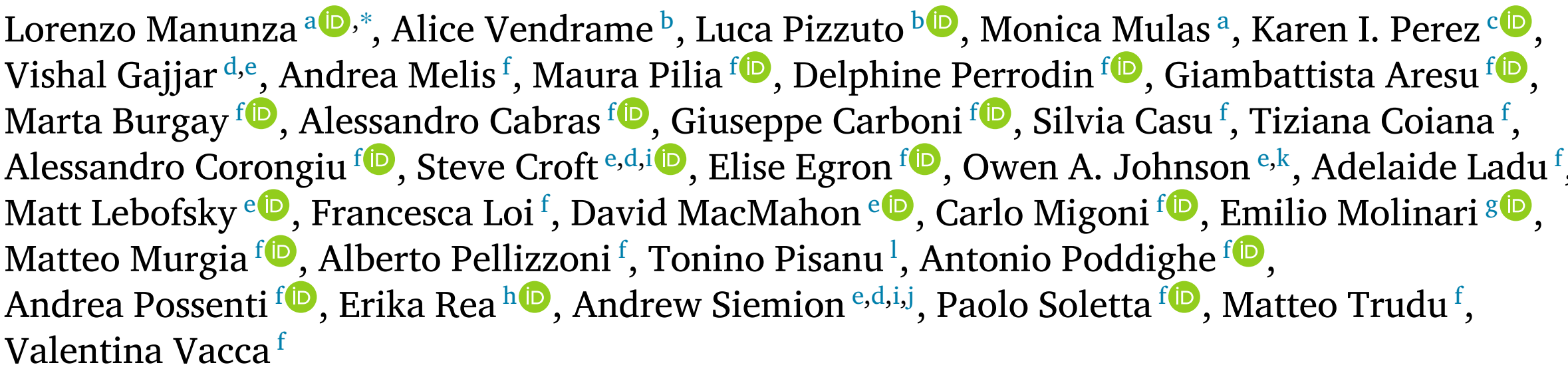

Lorenzo Manunza [a],*, Alice Vendrame [b], Luca Pizzuto [b], Monica Mulas [a], Karen I. Perez [c], Vishal Gajjar [d,e], Andrea Melis [f], Maura Pilia [f], Delphine Perrodin [f], Giambattista Aresu [f], Marta Burgay [f], Alessandro Cabras [f], Giuseppe Carboni [f], Silvia Casu [f], Tiziana Coiana [f], Alessandro Corongiu [f], Steve Croft [e,d,i], Elise Egron [f], Owen A. Johnson [e,k], Adelaide Ladu [f], Matt Lebofsky [e], Francesca Loi [f], David MacMahon [e], Carlo Migoni [f], Emilio Molinari [g], Matteo Murgia [f], Alberto Pellizzoni [f], Tonino Pisanu [l], Antonio Poddighe [f], Andrea Possenti [f], Erika Rea [h], Andrew Siemion [e,d,i,j], Paolo Soletta [f], Matteo Trudu [f], Valentina Vacca [f]

[a] *University of Cagliari, Department of Physics, Cittadella Universitaria di Monserrato (Building B), Italy*
[b] *ALMA MATER STUDIORUM - University of Bologna, Via Zamboni, 33, Bologna, Italy*
[c] *Department of Astronomy, Columbia University, 550 West 120th Street, NY, USA*
[d] *SETI Institute, 339 Bernardo Ave, Suite 200 Mountain View, CA 94043, USA*
[e] *Berkeley SETI Research Center, University of California, Berkeley, CA 94720, USA*
[f] *INAF – Osservatorio Astronomico di Cagliari, Via della Scienza 5, Selargius (CA), Italy*
[g] *INAF – Osservatorio Astronomico di Brera, Via Brera 28, 20121 Milan, Italy*
[h] *Sapienza University of Rome, Piazzale Aldo Moro 5, Rome, Italy*
[i] *Breakthrough Listen, University of Oxford, Department of Physics, Denys Wilkinson Building, Keble Road, Oxford, OX1 3RH, UK*
[j] *Institute of Space Sciences and Astronomy, University of Malta, Msida MSD2080, Malta*
[k] *School of Physics, Trinity College Dublin, College Green, Dublin 2, Ireland*
[l] *INAF – Istituto di Astrofisica e Planetologia Spaziali, Via del Fosso del Cavaliere 100, 00133 Rome, Italy*

## ARTICLE INFO

## ABSTRACT

The search for radio signals from technologically advanced extraterrestrial intelligence has traditionally focused around 1.4 GHz. In this paper, we extend the exploration into previously uncharted frequency space, presenting extensive observations at 6 GHz and pioneering the first survey at 18 GHz with the Sardinia Radio Telescope (SRT). Our strategy entailed rigorous observation sessions, totaling 73 h, directed towards the Galactic Center and 72 selected TESS targets—making this the most comprehensive high-frequency technosignature search to date. Our narrowband signal search found no definitive evidence of drifting signals that could suggest an extraterrestrial origin from the surveyed regions. Nevertheless, our efforts have enabled us to set new constraints on the presence of radio emissions from approximately $5 \times 10^5$ stars, establishing an isotropic radiated power limit of $1.8 \times 10^{19}$ W. We also provide a comparative analysis of the 'hits' recorded across both frequencies to highlight the significance of pursuing technosignature searches at higher frequencies, where the spectral landscape is less congested and more conducive to detection.

## 1. Introduction

The search for life in the Universe has been significantly enhanced by new discoveries of exoplanetary systems and planets with potentially habitable conditions outside the Solar System. However, due to the limitations of atmospheric characterization for such distant systems, any detected biosignatures could also result from abiotic processes [1] and may not provide conclusive evidence of life. Technosignatures, which are direct or indirect evidence of past or present technology developed by technologically-advanced extraterrestrial life, offer a clearer form of evidence for life. More specifically, we target narrowband drifting signals for two main reasons. First, as proposed by [2], narrowband signals (spectrally confined to around 1 Hz) are prime candidates for extraterrestrial beacons, as they are easy to generate and can propagate uninterrupted across vast interstellar distances. Furthermore, such narrowband signals do not naturally arise from any known astrophysical sources, eliminating the possibility of their origin being due to natural processes (see Enriquez et al. [3]) for detailed justifications).

* Corresponding author.
*E-mail address:* lorenzo.manunza99@gmail.com (L. Manunza).

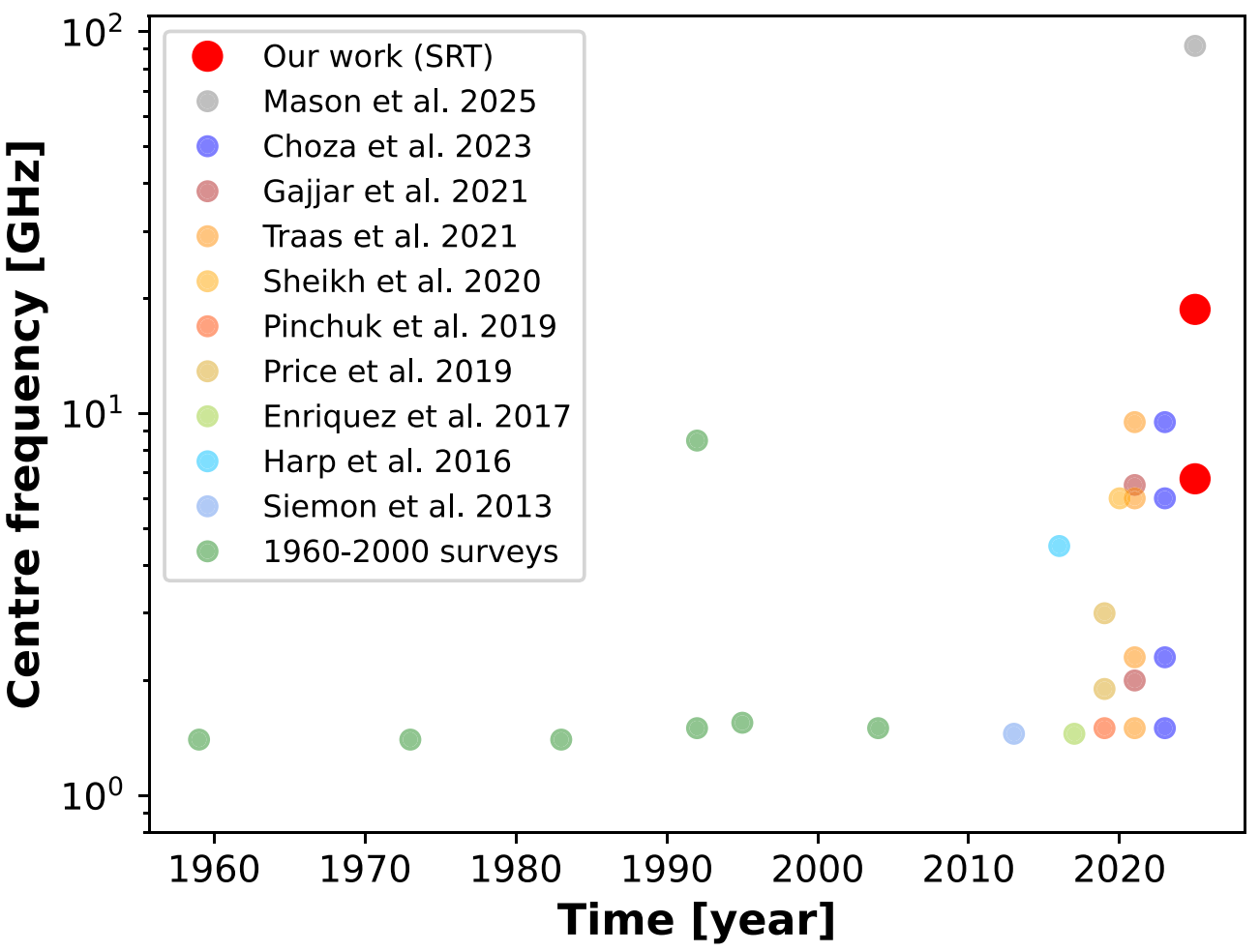

**Fig. 1.** Center frequency of several previous SETI surveys over the years. Historically, most of the earlier SETI surveys have been conducted around 1.4 GHz. Recent surveys have increasingly targeted higher frequencies, as this is essential for an agnostic search that does not assume the progenitor's choice of transmission frequency. The survey reported in this work is one of the very few high-frequency surveys ever conducted for technosignatures using ground-based radio telescopes.

Discovering a technosignature would provide more definitive proof of extraterrestrial life. To this end, modern surveys are actively searching for technosignatures [4]; low-frequency electromagnetic waves, such as coherent radio waves, are prime candidates for such beacons because they are energetically efficient to produce and can convey information at maximum speeds across interstellar distances.

An ideal technosignature search should be agnostic regarding the progenitor's choice of transmission frequency. In other words, we should not make any prior assumptions about the frequencies at which potential radio technosignatures could be produced by technologically advanced extraterrestrial life. Frequencies around 1 GHz were often searched, based on the argument presented by [2] regarding the significance of the hydrogen line. In Fig. 1, we list some of the most prominent radio technosignature searches since the 1960s. It is evident from Fig. 1 that many modern-day searches have also been conducted near 1 GHz due to the wide availability of receivers across a large number of radio telescopes. Higher-frequency receivers require greater design accuracy due to the smaller wavelengths involved, making them more economically expensive. Consequently, not all radio telescopes in the world are capable of observing at frequencies above 10 GHz. Furthermore, not many of these radio telescopes are involved in technosignature searches. There are multiple challenges associated with higher-frequency observations. For example, maintaining adequate surface accuracy becomes increasingly critical, and any surface degradation can lead to significant sensitivity loss. Hence, a large fraction of previous radio technosignature surveys have been carried out at frequencies below 10 GHz. However, there have been a handful of recent studies at higher frequencies [5,6]. Steffes et al. [7] carried out searches at 203 GHz towards 40 sun-like stars. However, their instantaneous bandwidth was only limited to 5 MHz which is significantly smaller compared to the bandwidth covered by most of the modern-day searches. Similarly, Shirai et al. [8] also conducted searches at 22 GHz, but only across 1.5 MHz of bandwidth with a spectral resolution of 6 kHz. Most recently, Mason et al. [9] reported one of the first high-frequency surveys using the Atacama Millimeter/Submillimeter Array at 90 GHz. Although this study did look for narrowband signals, the archival data products only allowed for searching in 30 kHz-wide channels. Thus, we can still argue that there is a substantial need to undertake technosignature searches at higher radio frequencies for truly narrowband signals ($\ll$1 Hz), covering the entire radio window accessible from ground-based facilities.

The Breakthrough Listen (BL) Program is the most comprehensive search for extraterrestrial life ever conducted, which aims to cover a large fraction of the parameter space. Within the BL program, searches are being carried out across a broad range of the electromagnetic spectrum, primarily using ground-based facilities. These searches target a wide range of objects, including the nearest million stars, the entire Galactic plane and center of the Milky Way, and a hundred nearby galaxies [10,11]. In 2019, the BL program began a new collaboration with Italy's Istituto Nazionale di Astrofisica (INAF) Cagliari Observatory to add the Sardinia Radio Telescope (SRT) to conduct searches at higher frequencies covering nearby stars and especially the Galactic Center region. The SRT is a 64-m single-dish radio telescope located in Sardinia, Italy, with state-of-the-art technological capabilities, allowing for high-frequency receivers up to 115 GHz [12].

In this work, we report one of the very first high-frequency surveys conducted towards a set of nearby stars and the Galactic Center. This is also one of the first surveys conducted and reported using the SRT, highlighting the growing number of radio telescopes joining the effort to expand the technosignature search to higher frequencies. In Section 1.1, we highlight our target selection process, and in Section 2, we discuss the details of the observations carried out using the SRT. In Section 3, we outline our signal search process and present our overall findings. In Section 4, we discuss the impact of our survey, including some interesting candidates we identified, and provide our final conclusions in Section 5.

### 1.1. Target selection

*TESS TOIs.* More than 5000 exoplanets have been confirmed, with over 800 systems having more than one planet. These have been primarily discovered using the transit method by various missions, such as NASA's Transiting Exoplanet Survey Satellite (TESS) and Kepler Mission. The significance of these missions extends to on-going technosignature searches. The ecliptic plane of the Solar System, where Earth resides, is likely to exhibit intense radiation from our planetary radar systems ($\geq 10^{13}$ W), which are routinely employed for studying Solar System bodies and monitoring potentially hazardous asteroids. It is expected that advancements in Solar System exploration will further increase this radiation leakage, particularly within the ecliptic plane. By analogy, the chances of detecting leakage radio emissions from ETIs on transiting exoplanets – which also reside in their own ecliptic planes – become significantly higher. The TESS mission [13] is designed to detect transiting planets around the nearest, brightest stars. Furthermore, compared to more distant exoplanets discovered by Kepler, exoplanets discovered by TESS are especially promising targets for technosignature searches due to their proximity. Objects listed in the TESS Input Catalog (TIC), which is compiled from various existing catalogs as detailed by [14], are monitored by TESS. Those that exhibit indications of transiting exoplanets after removing false positives through the deployed pipeline are then cataloged as TESS Objects of Interest (TOI)—a compilation of TIC subjects designated for further observational study. We meticulously reviewed the list of TOIs accessible as of 2019. From this, we compiled a list of approximately 1200 TOIs that are observable from the SRT. The rationale behind curating an extensive list is to ensure the availability of targets spanning the entire range of Local Sidereal Time (LST). See Fig. 2 for the distribution of these targets. In this work, we report 72 observations carried with the SRT towards TESS TOIs; this includes 42 targets observed at 6.5 GHz and 30 targets observed at 18 GHz, with 15 targets observed at both bands.

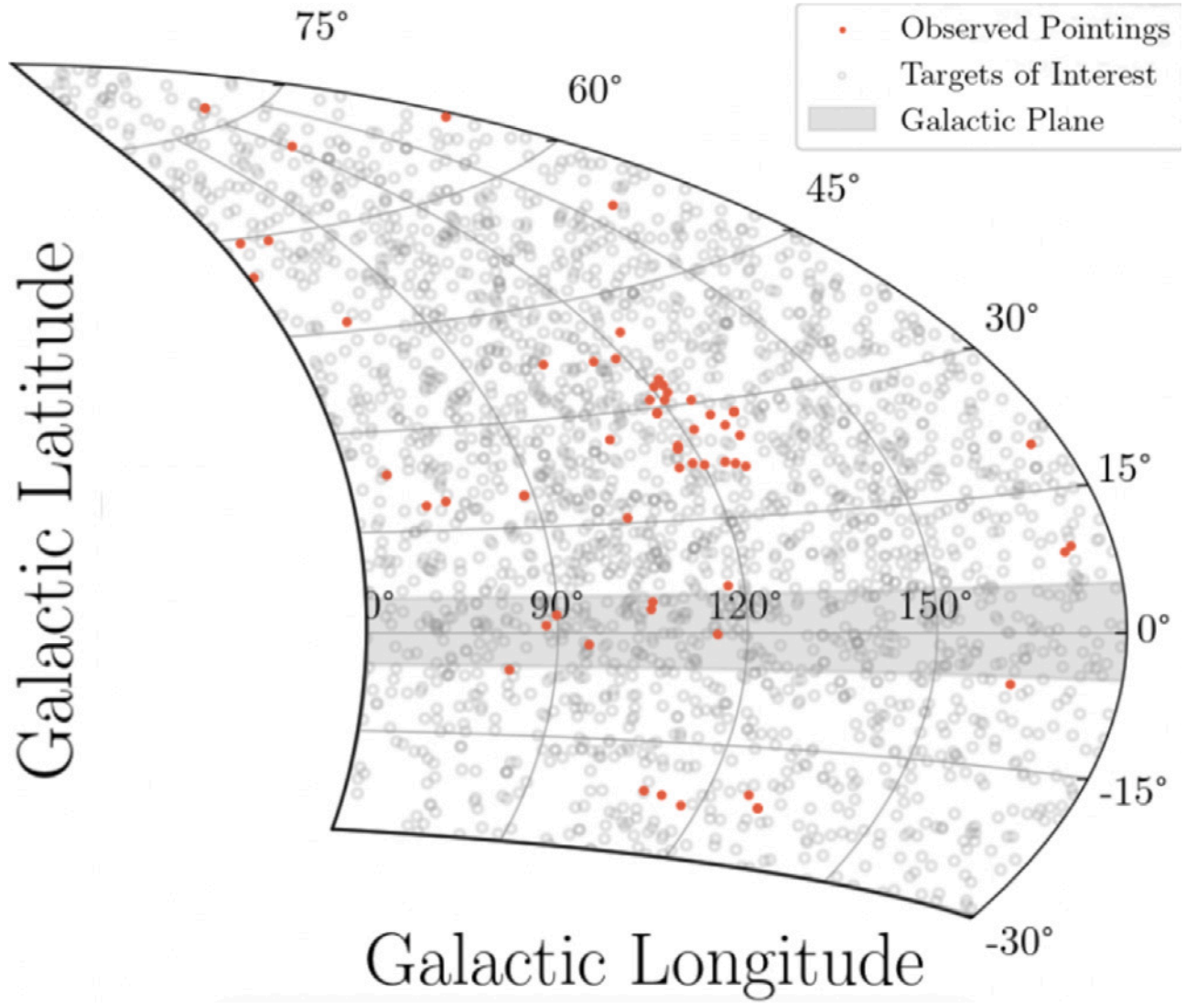

**Fig. 2.** Distribution of SRT targets from this work in galactic latitude (LAT) and galactic longitude(LON). The TESS project uses an all-sky survey approach to find transiting exoplanets.

Our observations were conducted on a dynamic schedule, allowing us to select targets in real-time based on their LST coverage for a given allotted observing session. For each session, we prioritized targets based on their distance. We plan to continue following up on the remaining targets in future observations to complete our survey of all 1200 TOIs. The sky distribution of the targets surveyed in this work is shown in Fig. 2. Table 1 presents a truncated list highlighting key features of the TESS targets we observed (the full list is available in the online version).

*Galactic center:*. The Galactic Center (GC) possesses the greatest concentration and highest number density of stars within the Galaxy. The line of sight towards the GC provides the most substantial integrated Galactic star count in comparison to any other direction in the sky. [15] presented three primary justifications for undertaking in-depth technosignature searches in the GC region. These justifications encompass the high likelihood of habitable planets and technologically advanced civilizations, and the potential for the GC to serve as a Schelling point for positioning a powerful transmitter with the capability to signal across the entire Milky Way. For our current work, we choose an approximately $9' \times 9'$ region around the GC to be surveyed at 6.5 GHz and an approximately $5' \times 5'$ region to be surveyed at 18 GHz. Each of these regions were divided into 19 and 37 individual pointings corresponding to the size of the Full Width Half Max (FWHM) beam widths of the SRT at 6.5 GHz and 18 GHz, respectively. This is similar to the region surveyed by the Green Bank Telescope and reported by [15]. Fig. 3, shows the surveyed region reported in this work at both these bands. The central pointing was labeled as A00 and the surrounding hexagon pointings as B, C, and D, numbering the beams counterclockwise starting immediately to the right of A00 for both bands.

## 2. Observations

### 2.1. Sardinia radio telescope

The SRT [16,17] is a 64-m single dish telescope located in Sardinia, Italy. It is among the most sensitive radio telescopes, equipped with receivers capable of conducting observations across a wide frequency range from 0.3 to 116 GHz [18]. Among different available backends, the SArdinia Roach2-based Digital Architecture for Radio Astronomy (SARDARA; [12]) employs flexible digital hardware that allows capturing and processing an instantaneous bandwidth of up to 28 GHz – divided into fourteen chunks each 2 GHz wide – from any of the selected receivers.We used SARDARA in a specialized mode, designed by the Breakthrough Listen team for the Green Bank Telescope (section 5 of [19]). This configuration was chosen for its efficient handling of raw channelized data and its ability to extract a 375 MHz bandwidth (the maximum the SRT's BL cluster supports) from SARDARA's 1500 MHz capacity. (For further details see Melis et al. [12]).

The SARDARA backend consists of seven ROACH2 boards, one of which was used for our current observations. The BL program has deployed two dedicated compute servers in order to capture raw baseband voltages; further servers are planned to be installed so as to increase the instantaneous bandwidth for future technosignature searches with the SRT. In our specific SETI configuration, we use an FPGA firmware with which the ROACH2 board handle, simultaneously, two input signals with a bandwidth of 1.5 GHz each (for a total of 3 GHz) and split them into 512 sub-bands, thereafter sent out via up to eight 10G SFP+ connections per ROACH2. Even though we could get to have the aforementioned 2 GHz for each IF, in order to avoid any packet loss in the packet capture framework, it has been preferred to reduce the input bandwidth to 1.5 GHz. Since the first 10G link is already used for the SRT conventional scientific observations, actually only seven 10G links are available for the BL backend, which means that the usable bandwidth is reduced to 1312.5 MHz for each of the two polarizations provided by SRT receivers. In order to simplify the infrastructure, each of the two servers is in a direct connection with the ROACH2 and records 1/8th of the available 1.5 GHz incoming band for each of the two polarizations. Thus, an instantaneous bandwidth of 187.5 MHz dual-polarization can be recorded and post-processed in real time at each of the two compute nodes. In order to cover a wider bandwidth, multiple observations of the same target need to be carried out by moving the start frequency accordingly. For instance, the C-high band (also known as M-band) receiver works at a frequency range of 5.7 – 7.7 GHz; thus, to start to observe at 5.7 GHz (taking into account that we can only record data from the second 10G output), the receiver's local oscillator has to be set at 5512.5 GHz, namely 5700–187.5 MHz. As a consequence, the first session will cover the frequency interval 5700–6075 MHz, the second session (by setting the local oscillator at 5887.5 MHz) will cover the interval 6075–6450 MHz, etc. Here, we report 50 h and 23 h of observations conducted at the frequency ranges 5.7–7.7 GHz (C-band) and 18.1–18.4 GHz (K-band), respectively. We observed two sets of targets from the SRT, as mentioned in Section 1.1

### 2.2. Observations of TESS TOIs

A total of 72 TESS TOIs were observed across both bands, with 42 observed at the C-band and 30 at the K-band, 15 of which were observed at both bands. In order to remove Radio Frequency Interference (RFI), which is defined as unwanted signals coming from anthropogenic activities, we observed reference OFF-target position (see [3]). Each TOI was observed for 15 min, consisting of three 5 min ON-scans interspersed with three 5 min OFF-scans. The OFF-targets were selected as regions of the sky that are $\pm 5°$ in declination away from the TOI. This strategy allowed for the effective elimination of false positives found during the ON-scans due to terrestrial interference, similar to the approach used in previous radio technosignature searches with single-dish telescopes (see [3]). Table 1 presents a truncated list of TESS targets and their observation bands reported in this work. For each observation, we were only able to cover 375 MHz of total bandwidth spanning 6573.5 MHz–6948.5 MHz and 18098.5 MHz–18473.5 MHz.

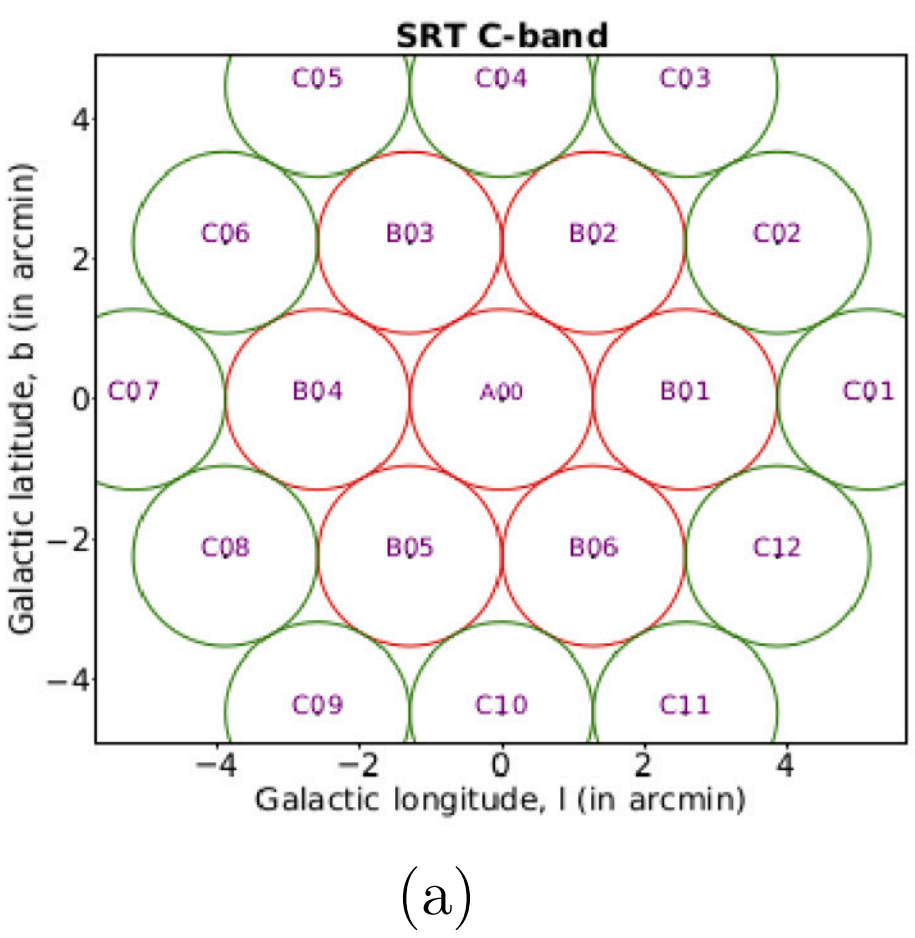

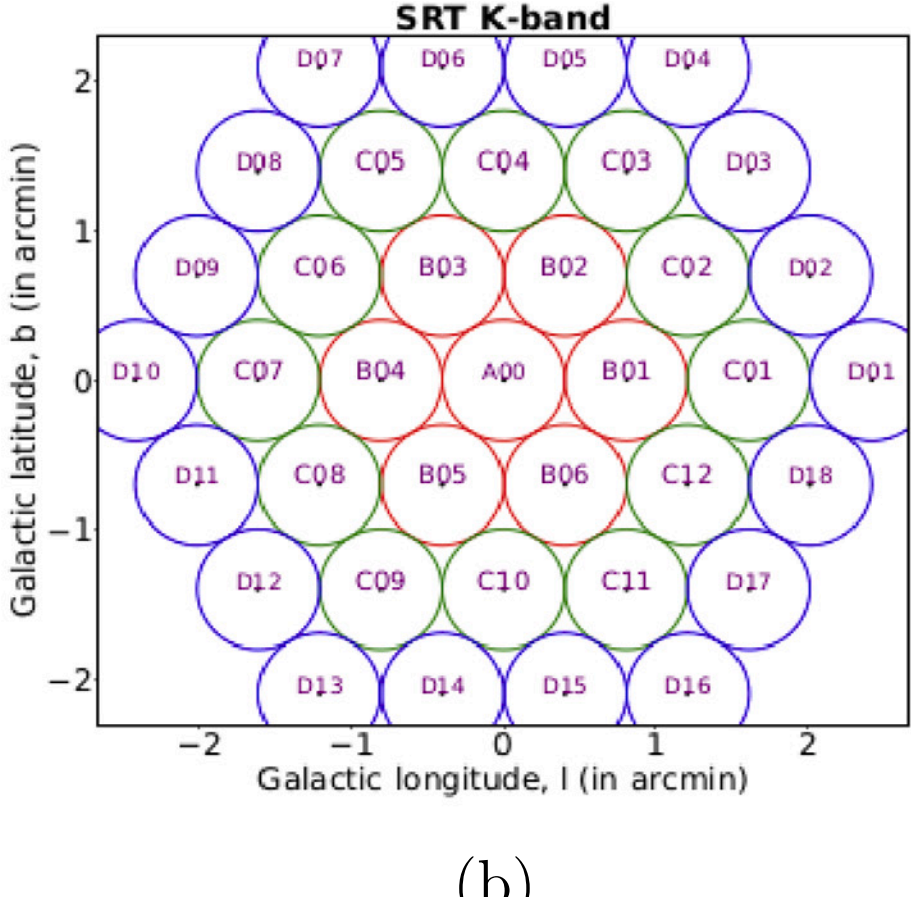

**Fig. 3.** Pointings for the proposed GC survey with the SRT. The central pointing, labeled A00, indicates the region of deep pointing of the GC(0, 0). **(a)** C band pointings to fully sample the $9' \times 9'$ GC bulge region. The Full Width Half Power beam width is 2.58' for each individual pointing **(b)** K band pointings to fully sample the $5' \times 5'$ GC bulge region. The Full Width Half Power beam width is 0.805' for each individual pointing.

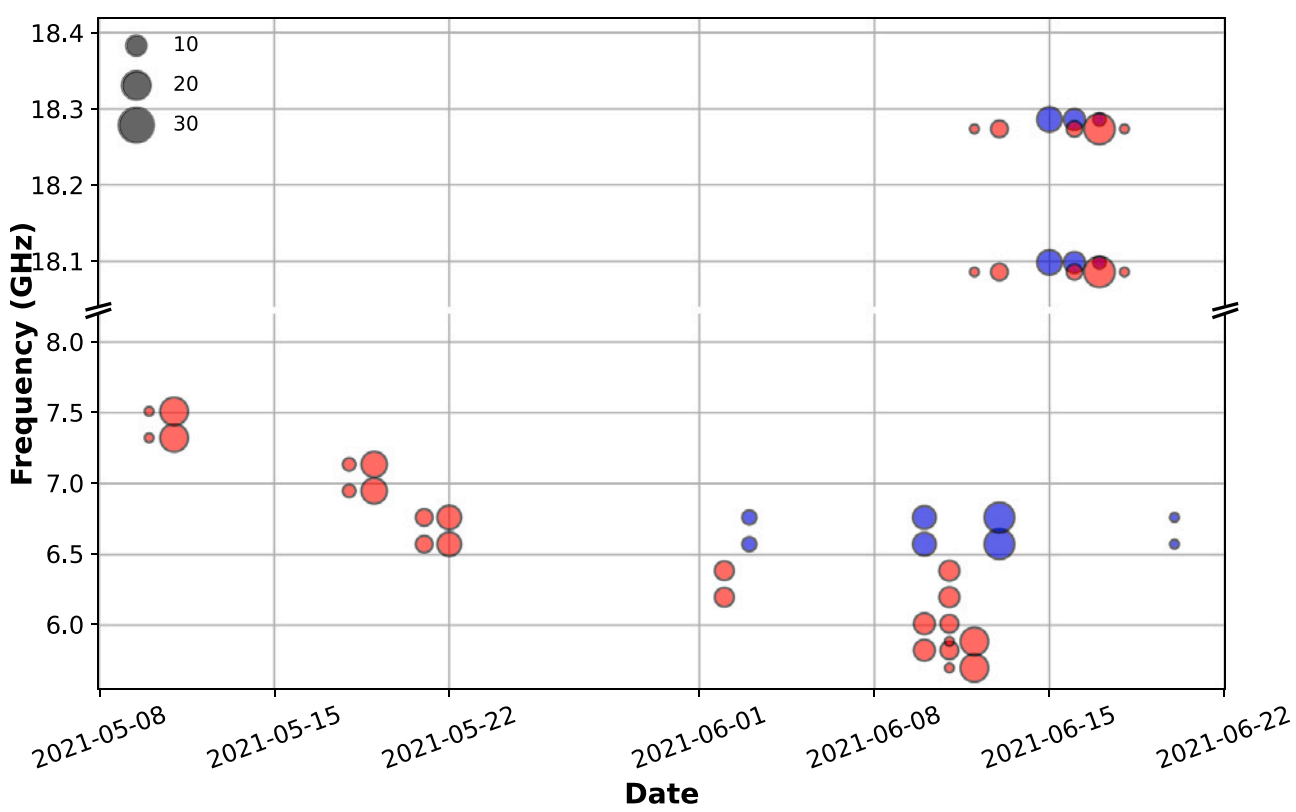

**Fig. 4.** Observations of two sets of targets that are reported in this work, along with their observing dates and central frequencies. The points in red represent observations of TESS TOIs, while the observations towards the GC pointings are marked in blue. The size of the markers provides a rough estimate of the number of targets observed on a given date. As can be seen, there are two pairs for each date, corresponding to the two compute nodes used to record 187.5 MHz of bandwidth per node (375 MHz combined from both nodes per observing session). To cover the 5.7 to 7.7 GHz, we switched frequencies, starting from the highest end and progressing to subsequent lower frequencies. (For interpretation of the references to color in this figure legend, the reader is referred to the web version of this article.)

### 2.3. Observations of GC region

To cover the GC region at both bands, as shown in Fig. 3, the region was divided into multiple pointings, with 19 pointings at C-band and 37 pointings at K-band. For both pointings, we labeled the central GC region A00. The surrounding hexagon pointings were labeled as B and C (plus D for K-band), with beams numbered counterclockwise starting immediately to the right of A00. To eliminate false positives in the narrowband searches, we conducted the observations in a sequence that allowed two FWHM separations between consecutive pointing centers. For example, in the C-band, short 5 min scans towards A00 were followed by shorter scans of outer ring pointings, such as C01 and C07. Similarly, other pointings were observed in pairs, such as B01–B04 and B02–B05. Tables 3 and 4 list all the pairs sequenced during our observations for both C-band and K-band, respectively. We repeated these observations for different frequency tuning during different observing sessions as shown in Fig. 4 to cover 5700 to 7700 MHz.

## 3. Analysis and results

During post-processing, a GPU-accelerated FFT tool was used on the recorded baseband voltages to produce high-spectral resolution products used to search for narrowband signals. From the recorded baseband voltages across each 187.5 MHz band, we produced SIGPROC-formatted filterbank products with 18.25 s and 2.79 MHz time and frequency resolutions, respectively, which were used for the remainder of the analysis. We report extensive searches for narrowband signals towards both sets of targets in Section 1.1. These signals are expected to drift in frequency due to Doppler shifts, originating from relative motion between the transmitter and receiver. We utilized the publicly available narrowband drifting signal search tool *turboSETI* [3] to conduct these searches.[1] For these searches, the signal-to-noise ratio (SNR) cut was kept at 10 with drift-rate ranges searched across ±4 Hz/s.

As described in Section 2, each target underwent three consecutive scans of 300 s, interspersed with 300 s of OFF-target scans to facilitate the removal of false positives. When observing the GC region, we used different pointings as ON and OFF pairs to better optimize on-source time. The initial run of *turboSETI* on each of these 300-s scans produced a total of 64,436 hits for TESS targets and 107,735 hits for GC across both bands. A "raw hit" is defined as a single strong narrowband signal in an observation that exceeds the set threshold in at least one fine channel. The distribution of these hits across both observation bands and all targets is detailed in Table 2. A hit present in the ON observations and absent in the OFF observations is considered an *event*. We take the list of raw hits produced for every pair of ON-OFF observations and compare them across the 6 segments (3 ONs and 3 OFFs). For the TESS TOIs, these OFFs are empty sky positions at ±5°, while for the GC pointings, we used different pointings as OFFs. We categorize these events into two groups as listed in Section 3. *Filter 2* events are hits that are present in at least one ON scan and absent in all three OFF scans. *Filter 3* events are hits that are present in all three ON scans and absent in all three OFF scans. A list of these events observed at each observing band is shown in Table 2 for both sets of targets. Tables 3 and 4 provide a detailed list of filtered events for each pair of the GC pointings at C-band and K-band, respectively.

The distribution of hits and events as a function of observed frequency is shown in Figs. 5 and 6 for TESS and GC targets, respectively. From Fig. 5, it is evident that the distribution of raw hits peaks near 6.75 GHz, which is likely due to radio frequency interference from

[1] For a more detailed discussion on *turboSETI* and its limitations, we recommend readers refer to [6,15].

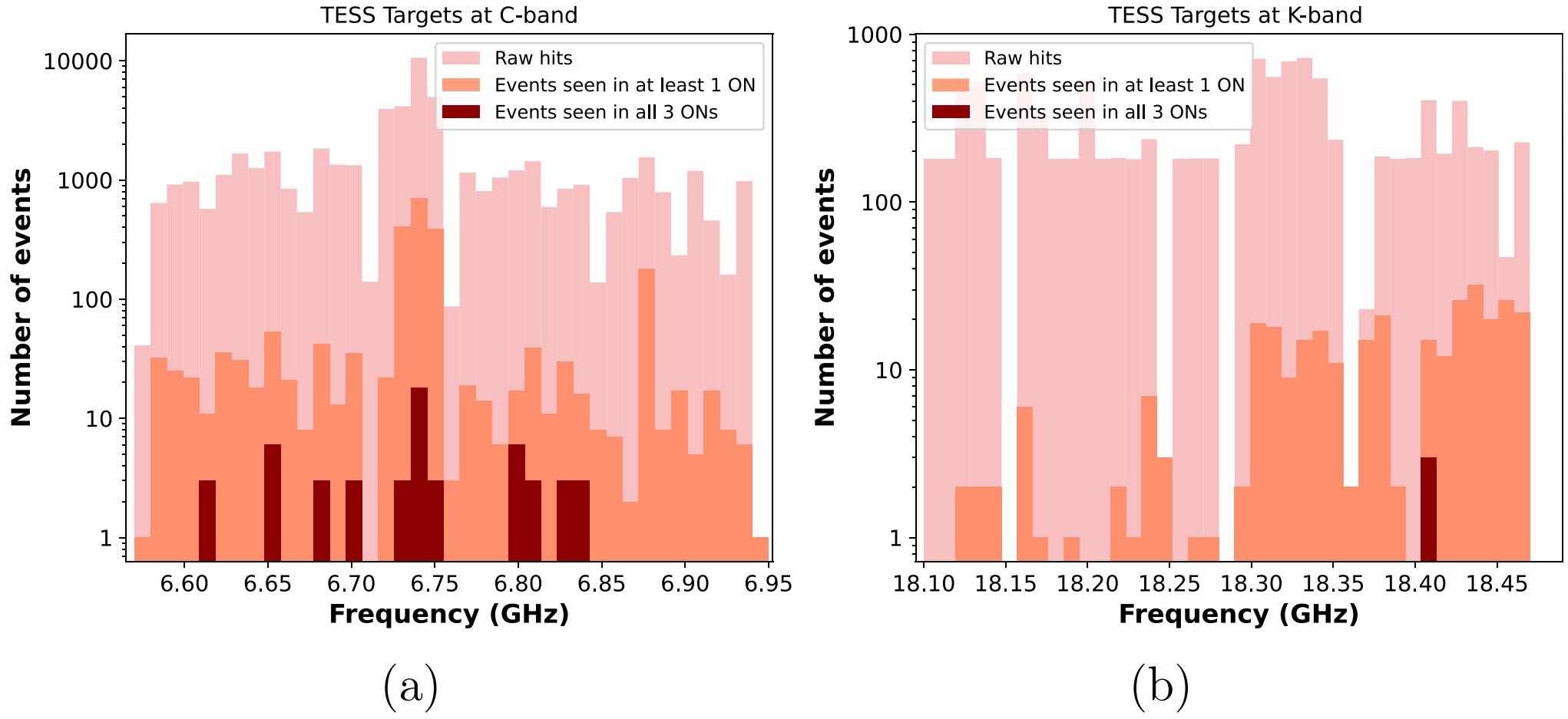

**Fig. 5.** Distribution of raw hits and subsequent filtered events obtained from the narrowband drifting search conducted towards TESS TOIs at (a) C-band and (b) K-band. For these sources only single 375 MHz tuning was used for observations. The histogram of hits at C-band shows an order of magnitude larger number of hits compared to raw hits recorded at K-band suggesting relatively interference free environment at higher frequencies.

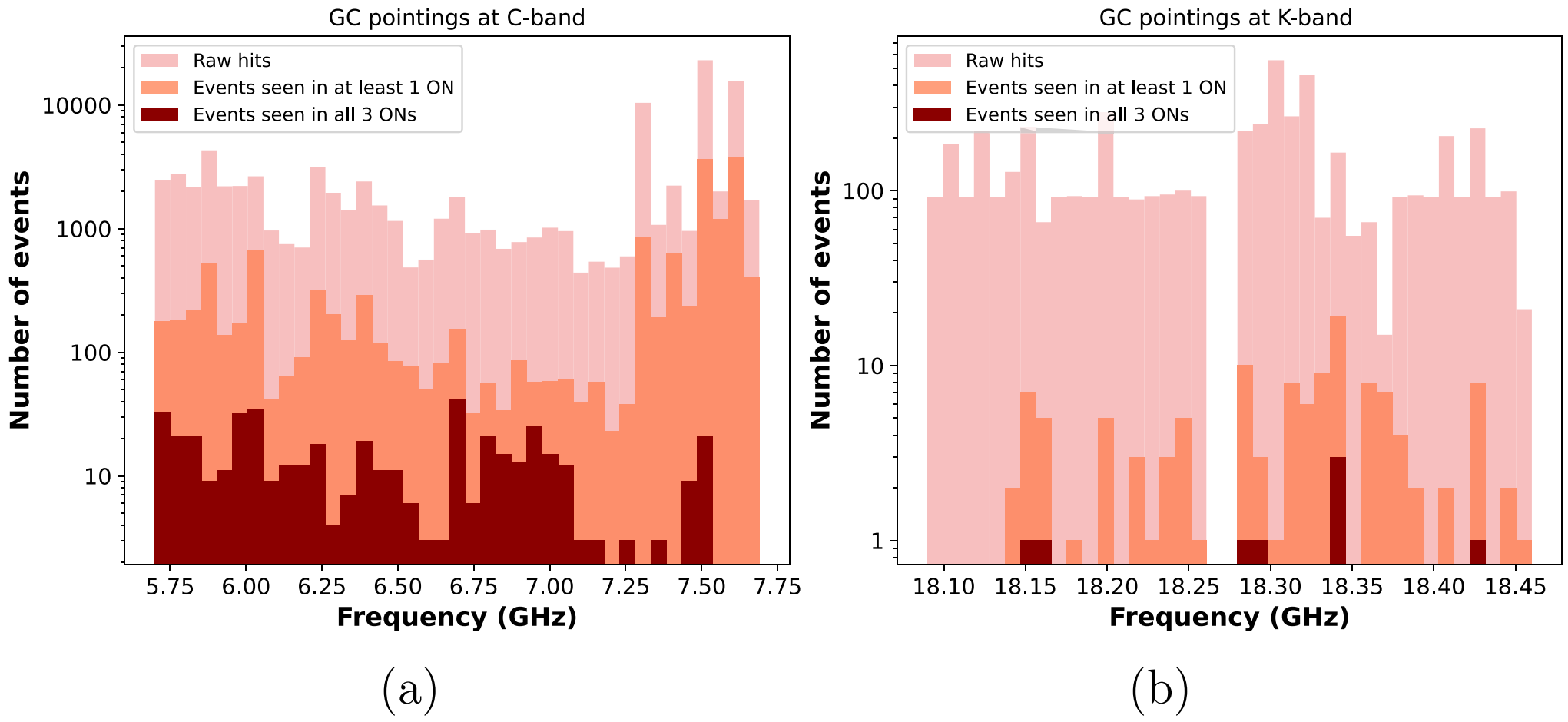

**Fig. 6.** Distribution of raw hits and subsequent filtered events obtained from the narrowband drifting search conducted towards GC pointings at (a) C-band and (b) K-band. For these GC pointings different tuning of 375 MHz was used on different sessions to cover the entire 5700 to 7700 MHz of bandwidth. For the K-band, we only used a single 375 MHz tuning for our observations.

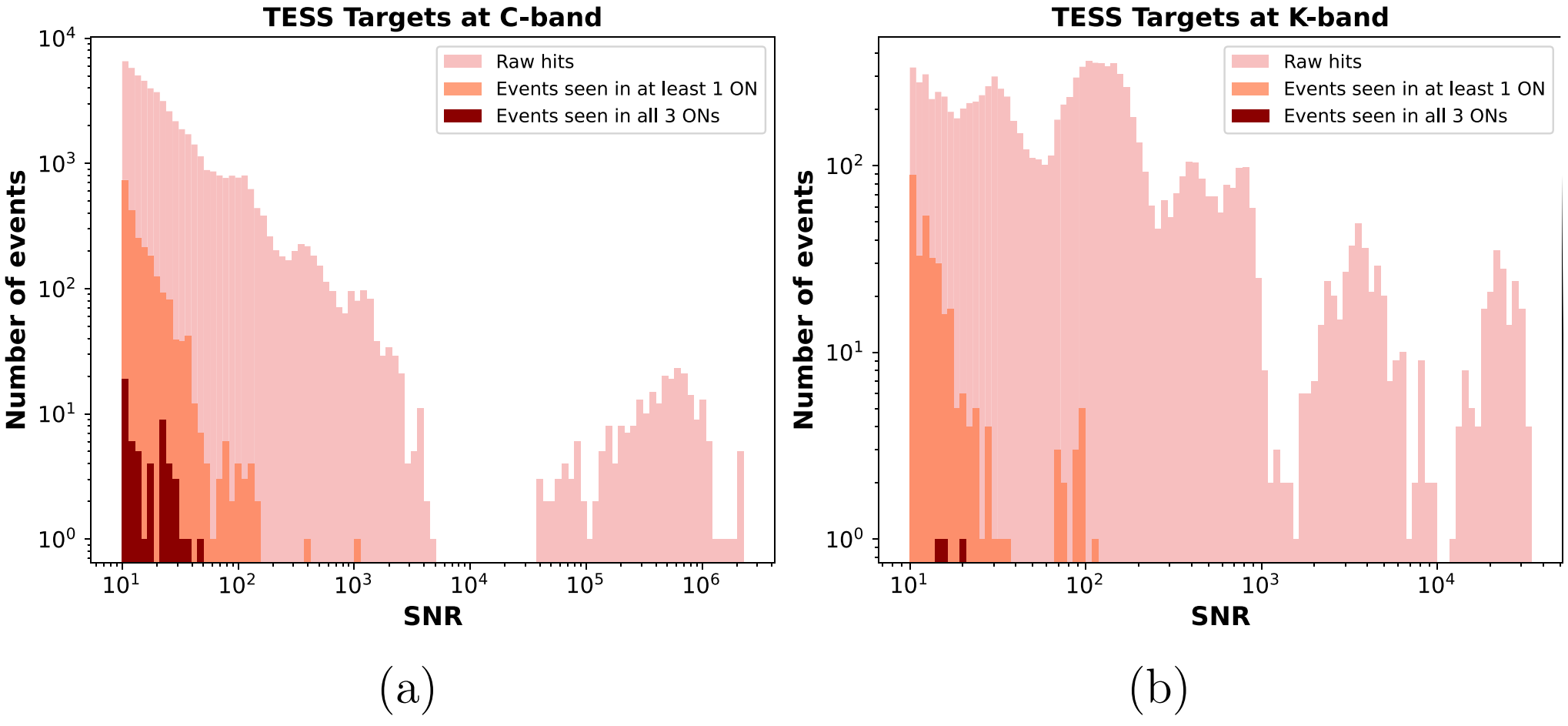

**Fig. 7.** SNR of detected hits and subsequent filtered events from the observations of the TESS TOIs at (a) C-band and (b) K-band.

**Table 1**
TESS TOIs observed. The columns, from left to right, are: TIC ID, observed frequency (C and K band), right ascension and declination, the date of the observation, the distance in pc, the spectral type of the star, the age of the star, the mass of the star and the luminosity of the star.

Truncated list of targets analyzed in this work's search for broadband pulsed SETI beacons

| TIC ID | Frequency range (GHz) | R.A. (HH:MM:SS) | Decl (HH:MM:SS) | Local date of observation (YYYY-MM-DD) | Distance (pc) | Spectral type | Age of the star (GYr) | Mass of the star ($M_{\odot}$) | Luminosity of the star ($L_{\odot}$) |
|---|---|---|---|---|---|---|---|---|---|
| 264678534 | 6.575–6.950 | 21:40:44.78 | 84:20:00.56 | 2021/06/13 | 28.25 | K0 V | | | |
| | 18.1–18.475 | 21:40:44.78 | 84:20:00.56 | 2021/06/16 | 28.25 | K0 V | $7.1^{-4.4}_{+4.5}$ | $0.84 \pm 0.05$ | $0.360^{-0.014}_{+0.019}$ |
| 27491137 | 18.1–18.475 | 14:29:34.24 | 39:47:25.54 | 2021/06/15 | 41.923 | G | $0.204^{-0.050}_{+0.053}$ | $0.850^{-0.026}_{+0.025}$ | $0.3777^{-0.0092}_{+0.0094}$ |
| 427730490 | 18.1–18.475 | 23:29:05.79 | 71:30:22.91 | 2021/06/16 | 144.717 | ... | ... | ... | 0.478 |
| 333473672 | 6.575–6.950 | 23:36:57.89 | 39:38:21.89 | 2021/06/13 | 154.56 | K0 V | | | |
| | 18.1–18.475 | 23:36:57.89 | 39:38:21.89 | 2021/06/15 | 154.56 | K0 V | $5.1^{-3.1}_{+3.9}$ | $0.981^{-0.065}_{+0.062}$ | $1.116 \pm 0.037$ |
| 154741689 | 6.575–6.950 | 10:57:14.53 | 89:05:13.12 | 2021/06/10 | 206.368 | ... | ... | ... | 1.435 |
| 432549364 | 6.575–6.950 | 00:01:26.9 | 39:23:01.66 | 2021/06/03 | 271.5 | F5 | ... | ... | 3.107 |
| ... | | | | | | | | | |
| 154872375 | 6.575–6.950 | 12:53:34.92 | 85:07:46.10 | 2021/06/13 | 113.92 | G0 D | ... | ... | 1.657 |
| | 18.1–18.475 | 12:53:34.92 | 85:07:46.10 | 2021/06/17 | 113.92 | G0 D | ... | ... | 1.657 |
| 154867950 | 6.575–6.950 | 12:33:26.52 | 84:40:44.93 | 2021/06/13 | 129.23 | G5 D | ... | ... | 2.817 |
| | 18.1–18.475 | 12:33:26.52 | 84:40:44.93 | 2021/06/17 | 129.23 | G5 D | ... | ... | 2.817 |
| 174302697 | 6.575–6.950 | 23:28:42.78 | 47:20:58.27 | 2021/06/13 | 168.87 | F2 D | ... | ... | 4.498 |
| | 18.1–18.475 | 23:28:42.78 | 47:20:58.27 | 2021/06/15 | 168.87 | F2 D | ... | ... | 4.498 |
| 176899385 | 6.575–6.950 | 23:39:05.81 | 42:27:57.50 | 2021/06/13 | 272.68 | F8 C | ... | ... | 4.238 |
| | 18.1–18.475 | 23:39:05.81 | 42:27:57.50 | 2021/06/15 | 272.68 | F8 C | ... | ... | 4.238 |

**Table 2**
Number of hits and filter 2 and 3 events, categorized by sub-band as discussed in Section 2. Most pairs for each sub-band had a complete cadence of six scans, except for pairs B03–B06 and C03–C05. In sub-bands 5823.5–6011.0 and 6011.0–6198.5, the B03–B06 pair had only four scans (2 per pointing). For sub-bands 6198.5–6386.0 and 6386.0–6573.5, the B03–B06 pair had just two scans (1 per pointing). In sub-bands 6573.5–6761.0 and 6761.0–6948.5, the same pair had three scans (two and 1 per pointing, respectively).

Distribution of hits and events per sub-band

| | Frequency (MHz) | Total number of hits | Events passing filter 2 | Events passing filter 3 |
|---|---|---|---|---|
| TESS targets | 6573.5–6761.0 | 38 695 | 1865 | 39 |
| | 6761.0–6948.5 | 15 077 | 412 | 15 |
| | 18 098.5–18 286.0 | 4711 | 29 | 0 |
| | 18 286.0–18 473.5 | 5953 | 284 | 3 |
| Galactic center | 5698.5–5886.0 | 9481 | 511 | 57 |
| | 5823.5–6011.0 | 6683 | 432 | 55 |
| | 5886.0–6073.5 | 6646 | 837 | 27 |
| | 6011.0–6198.5 | 4724 | 479 | 50 |
| | 6198.5–6386.0 | 8149 | 857 | 52 |
| | 6386.0–6573.5 | 4336 | 394 | 30 |
| | 6573.5–6761.0 | 4153 | 279 | 50 |
| | 6761.0–6948.5 | 3006 | 220 | 53 |
| | 6948.5–7136.0 | 3128 | 214 | 57 |
| | 7136.0–7323.5 | 5299 | 673 | 3 |
| | 7323.5–7511.0 | 33 674 | 4925 | 33 |
| | 7511.0–7698.5 | 19 685 | 5435 | 0 |
| | 18 086.0–18 273.5 | 2224 | 33 | 2 |
| | 18 273.5–18 461.0 | 3127 | 90 | 6 |

satellite downlink and TV broadcasts known to operate in the C-band. At K-band, the raw hits distribution is relatively flat, and the number of hits is an order of magnitude lower compared to the number of raw hits detected at C-band. Figs. 7 and 8 show the distribution of observed SNR of raw hits and events at C-band and K-band for all our targets. It can be seen that C-band exhibits a distribution with a large number of very high SNR raw hits compared to K-band. The peaks in these distributions at higher SNR were all detected with near-zero drift. These are likely the DC spikes from the FFT applied during post-processing to produce the fine channelized data products.

By default, *turboSETI* eliminates hits produced by DC spikes. However, a few are missed and they appear as near-zero-drifting signals across the ON-OFF observing pairs, clearly indicating they are terrestrial. Figs. 9 and 10 show the distribution of drift rates detected at C-band and K-band for TESS TOIs and GC pointings, respectively. An interesting observation can be made when comparing the distributions of detected drift rates between the C-band and K-band. The distribution of drift rates at C-band appears to extend beyond the survey parameter limit of ±4 Hz/s, while almost all of the hits detected at K-band only span between ±0.4 Hz/s. We examined this behavior in detail by analyzing the distribution of hits as a function of frequency channels. At C-band, we identified approximately 0.08 MHz across 2000 MHz exhibiting strong RFI (0.004% of the total band), which were responsible for nearly all high drift-rate hits. Removing these RFI-affected channels reduced the number of high drift-rate signals at C-band by about an order of magnitude, bringing its drift rate distribution in line with that of K-band, both showing a spread within ±1 Hz/s. Additionally, to ensure that our pipeline was not filtering out high-drift rate signals at K-band, we conducted a validation test by injecting simulated signals with varying drift rates into our data. We then performed a blind detection of these injected signals and found no significant difference in detection efficiency across different drift rates. This confirms that the absence of high drift-rate hits at K-band is not due to a bias in our pipeline but rather a real effect, likely driven by RFI differences between the two bands.

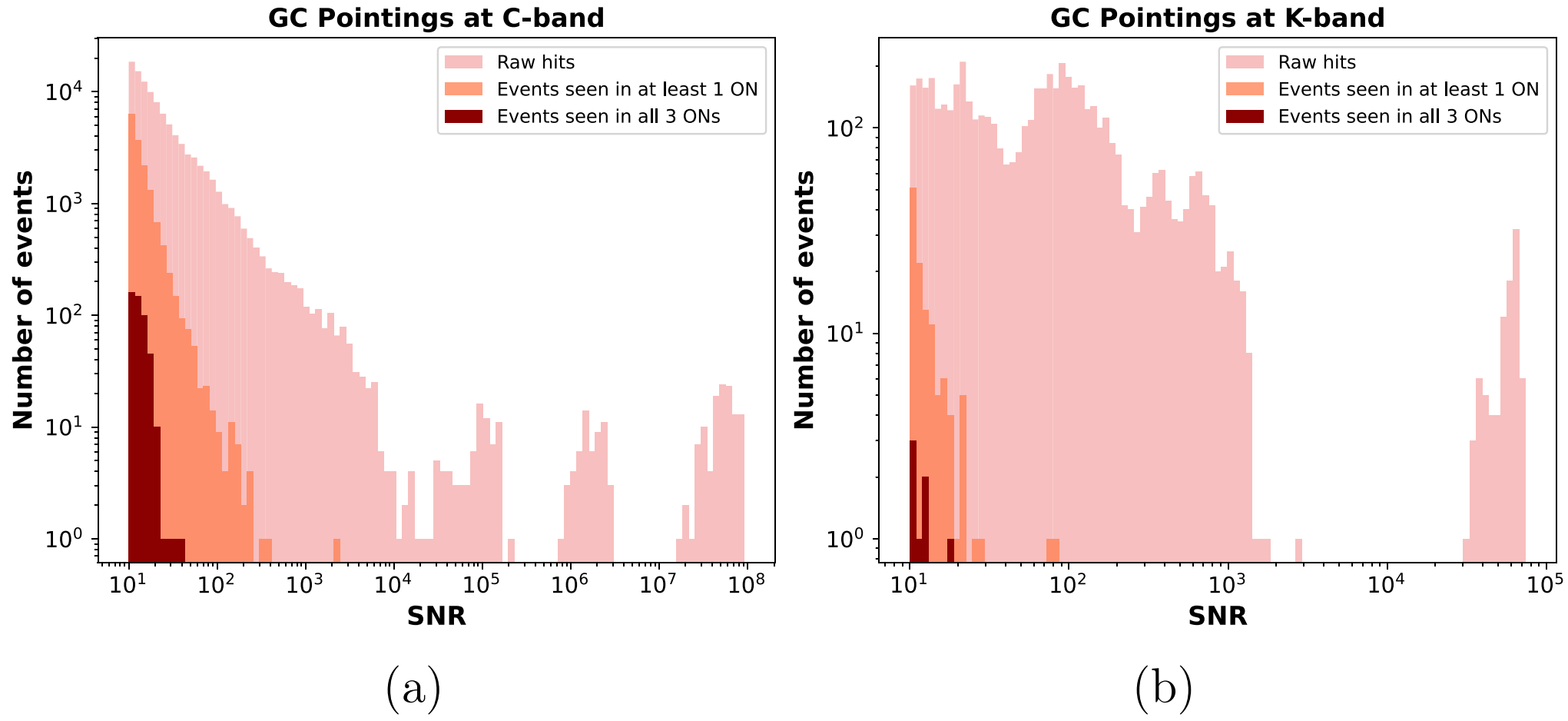

**Fig. 8.** SNR of detected hits and subsequent filtered events from the observations of the GC pointings at (a) C-band and (b) K-band.

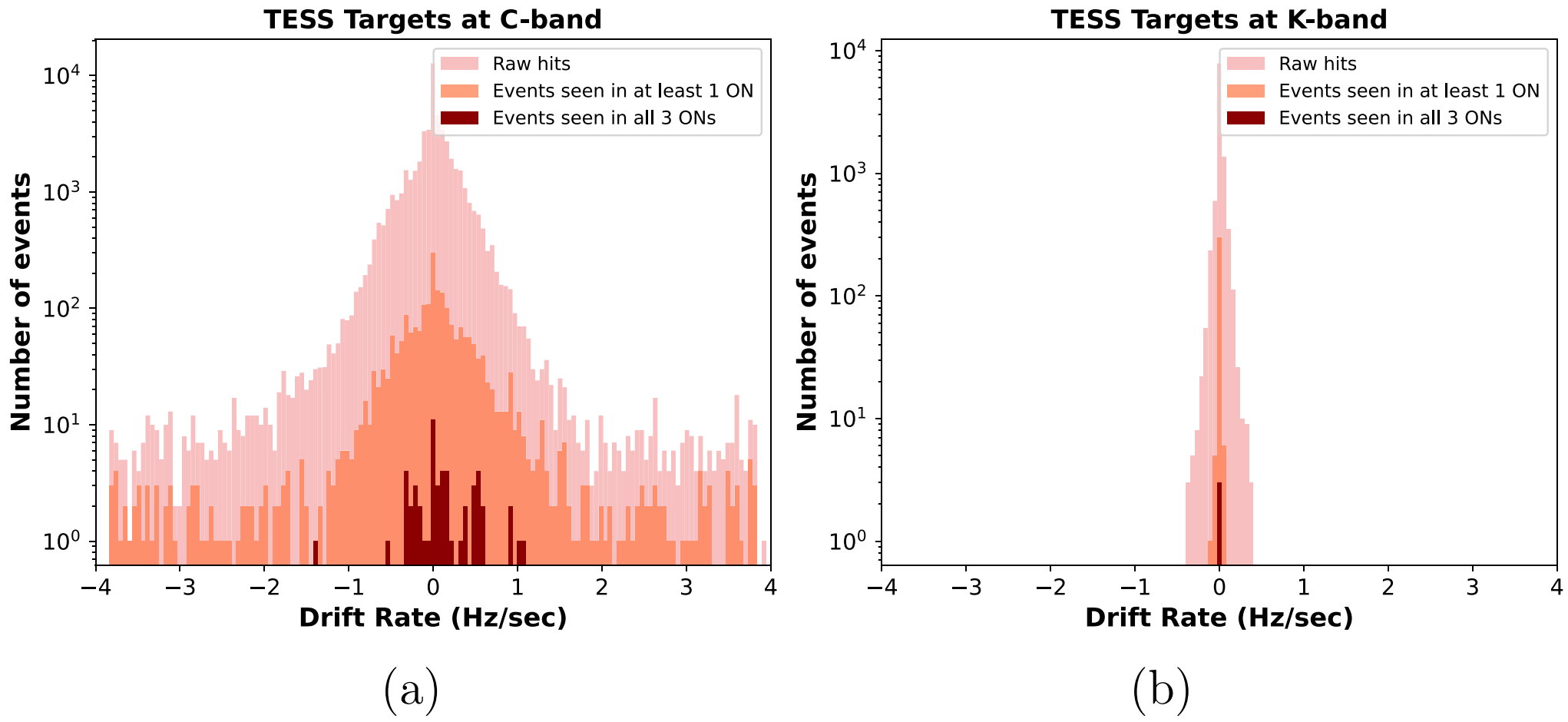

**Fig. 9.** Drift-rates of hits from TESS TOIs at (a) C-band and (b) K-band.

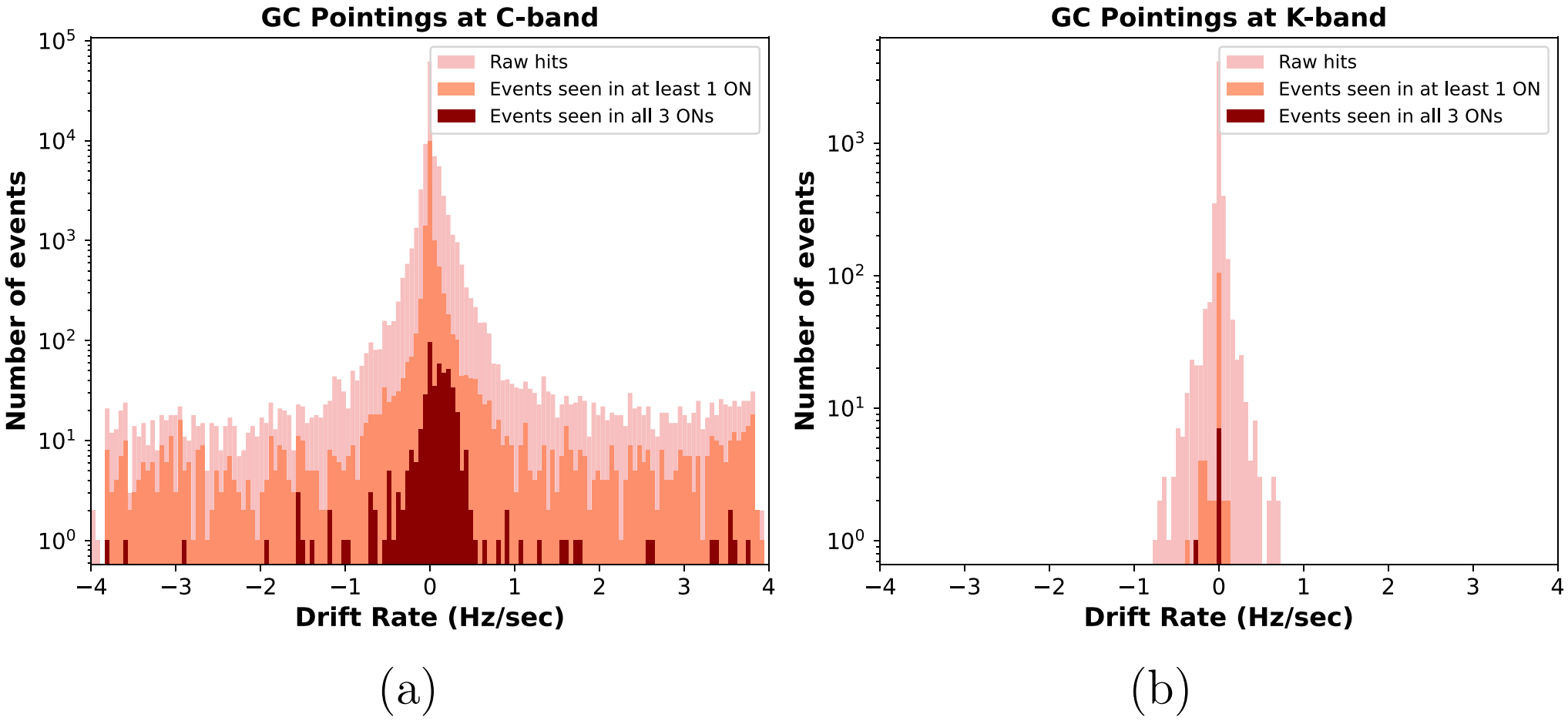

**Fig. 10.** Drift-rates of hits from GC pointings at (a) C-band and (b) K-band.

## 4. Discussion

### *4.1. Candidates*

For the final verification, we plotted all the filter 2 and filter 3 events for both sets of targets at both bands and visually examined them. For the GC region observations at K-band, we observed only around 8 events at filter 3 combined from all the pairs, as evident from Tables 2 and 4. These appear to occur only because we had some missing scans, preventing us from having the full 6 subsequent scans for a true comparison. For the TESS TOI observations at K-band, we also had a very small number of events at both C-band and K-band, and all of them appear to be due to incorrect identification of zero drift-rate signals between ON and OFFs. In all cases, we were able to identify narrowband signals across all the ONs and OFFs, which our automated pipeline missed due to incorrect identification of the drift rates and, in some cases, lower SNR of the narrowband signal present during the OFF scans. From our visual inspection of all the events at C-band, we did not find any convincing evidence of narrowband drifting signals at C-band for both sets of targets.

**Table 3**
Summary of BL-GC survey conducted with the SRT across 7.5–7.7 GHz. The right ascension, declination, and Galactic coordinates for each ON-source pointing are listed in columns 3–6. Columns 7 and 8 show the number of candidates detected by turboSETI for each ON-source, for filters 2 and 3. Most pointings had three scans, for a total cadence of six scans per pair, for each of the six sub-bands. There were two pairs that had exceptions, but because they are sub-band specific, they are detailed in Table 2.

| SRT C-band: 4 to 8 GHz | | | | | | | |
|---|---|---|---|---|---|---|---|
| Pair | ON-source pointing | RA (J2000) (h:m:s) | DEC (J2000) (d:m:s) | l (deg) | b (deg) | f2 Events | f3 Events |
| A00-C01 | A00 | 17:45:40.04 | −29:00:28.10 | 359.944 | −0.046 | 38 | 0 |
| | C01 | 17:45:52.33 | −28:56:03.79 | 0.03 | −0.046 | 1398 | 213 |
| C07-A00 | C07 | 17:45:27.74 | −29:04:52.34 | 359.858 | −0.046 | 4375 | 159 |
| | A00 | 17:45:40.04 | −29:00:28.10 | 359.944 | −0.046 | 194 | 0 |
| B01-B04 | B01 | 17:45:46.19 | −28:58:15.96 | 359.987 | −0.046 | 1426 | 0 |
| | B04 | 17:45:33.89 | −29:02:40.23 | 359.901 | −0.046 | 537 | 0 |
| B02-B05 | B02 | 17:45:34.39 | −28:58:12.19 | 359.966 | −0.009 | 664 | 6 |
| | B05 | 17:45:45.69 | −29:02:44.00 | 359.923 | −0.083 | 592 | 9 |
| B03-B06 | B03 | 17:45:28.24 | −29:00:24.30 | 359.923 | −0.009 | 344 | 16 |
| | B06 | 17:45:51.84 | −29:00:31.84 | 359.966 | −0.083 | 573 | 40 |
| C02-C04 | C02 | 17:45:40.54 | −28:56:00.06 | 0.009 | −0.009 | 748 | 0 |
| | C04 | 17:45:22.60 | −28:58:08.35 | 359.944 | 0.028 | 807 | 0 |
| C03-C05 | C03 | 17:45:28.75 | −28:56:56.26 | 359.987 | 0.028 | 1031 | 8 |
| | C05 | 17:45:16.45 | −29:00:20.43 | 359.901 | 0.028 | 299 | 7 |
| C08-C06 | C08 | 17:45:39.54 | −29:04:56.14 | 359.88 | −0.083 | 351 | 3 |
| | C06 | 17:45:22.09 | −29:02:36.39 | 359.88 | −0.009 | 354 | 0 |
| C11-C09 | C11 | 17:46:03.63 | −29:00:35.51 | 359.987 | −0.121 | 381 | 3 |
| | C09 | 17:45:51.35 | −29:04:59.88 | 359.901 | −0.121 | 363 | 0 |
| C10-C12 | C10 | 17:45:57.49 | −29:02:47.71 | 359.944 | −0.121 | 331 | 3 |
| | C12 | 17:45:57.98 | −28:58:19.66 | 0.009 | −0.083 | 450 | 0 |

Fig. 11 lists four interesting filter 2 events detected from our survey. Fig. 11(top left) is an interesting case where a narrowband signal appears to be seen drifting for a brief moment only during the observations of TESS TIC 432549364 at the frequency of 6934.311 MHz. Although, this signal does not appear to repeat during the OFF scans, we also did not detect it in other ON observations of the target which is common for spurious interference. Fig. 11(top right) shows an example of a filter 2 event from TIC 174302697 at 18329.389 MHz, as a representation of many such events seen at C-band and K-band. Here, a bright signal is present during the first ON observation, but during the subsequent observations, the signal is relatively weaker, yet present throughout the length of the observation. Such weaker signals sometimes get missed by the *turboSETI* pipeline due to the SNR threshold. Fig. 11(bottom left) shows an filter 2 event for the C03–C05 pairs, with C03 as the ON source. At first glance, it appears that the signal is only present during the first and third ONs, but upon closer examination, the signal can be seen towards the end of the second OFF scan. Fig. 11(bottom right) again shows a representation of a typical RFI event detected for the D08–D11 pair at 18433.316 MHz. We have noticed many such filter 2 and filter 3 events that can easily be mischaracterized due to changing drift rates. From all our visual inspections of all events, we did not find any convincing evidence of narrowband drifting signals at C-band nor K-band for either set of targets.

Previous experiments have demonstrated that *turboSETI* recovers 100% of injected signals with high S/N–even at high drift rates beyond its search bounds–albeit with inaccurate drift rate and central frequency estimates due to Doppler smearing (“dechirphing loss”) and significant S/N attenuation [6]. Although these tests were designed to quantify sensitivity loss rather than simulate weak signals, low-power signals or those embedded in noise could experience sufficient dechirping loss to fall below the detection thresholds, resulting in false negatives. Such limitation motivates our efforts to integrate machine learning approaches to improve detection sensitivity and reduce the risk of overlooking real signals [20].

### 4.2. Survey sensitivity

The sensitivity of narrowband ETI signal searches can be derived based on the following relation given by [15],

$$S_{min,narrow} = \frac{SNR_{min}}{\beta} \frac{S_{sys}}{\delta\nu_t} \sqrt{\frac{\delta\nu}{n_\rho \tau_{obs}}}. \tag{1}$$

Here, $S_{sys}$ is the system equivalent flux density, $\tau_{obs}$ is the total integration time for a given scan, $n_\rho$ is the number of polarizations, and $\delta$ is the bandwidth of the transmitted signal, which is assumed to be close to 1 Hz. Due to our very narrow frequency resolution (2.8 Hz) and relatively low temporal resolution (18.25 s), higher drifting signals get smeared across multiple frequency channels, thus reducing the detection sensitivity. This reduction in sensitivity can be addressed with the dechirping efficiency $\beta$, which depends on the frequency and time resolution of the spectra as reported by Gajjat et al. [15]. For the SRT, the minimum flux density can be derived through the radiometer equation, again following [15].

We can use the minimum detectable flux density given Eq. (1) to calculate the Equivalent Isotropic Radiated Power (EIRP) of a hypothetical ETI transmitter, located at distance $D$, that we can detect as:

$$EIRP = S_{min,narrow} \times 4\pi D^2. \tag{2}$$

For our survey of the GC, the EIRP sensitivity was $1.4 \times 10^{20}$ W at C-band and $8 \times 10^{19}$ W at K-band. For the survey of the TESS targets, the largest distance in our sample was around 272 pc, which we can also use in Eq. (2) to estimate an EIRP sensitivity of $3.7 \times 10^{16}$ W at C-band and $2 \times 10^{16}$ W at K-band[2]. For comparison, a Kardashev Type I civilization typically harnesses around $10^{16}$ to $10^{17}$ W, which is within the sensitivity range of our survey for nearby targets within 272 pc. In contrast, a Kardashev Type II civilization would harness approximately $10^{26}$ W, easily detectible by our survey for both C-band and K-band observations.

[2] For these calculations, we have assumed $\beta = 1$ for simplicity. The EIRP sensitivity will decrease by approximately a factor of $10^2$ for a signal drifting at the highest drift rate (±4 Hz/s) in our survey.

**Table 4**

Summary of BL-GC survey conducted with the SRT across 18–18.5 GHz. The right ascension, declination, and Galactic coordinates for each ON-source pointing are listed in columns 3–6. Columns 7 and 8 show the number of candidates detected by turboSETI for each ON-source, for filters 2 and 3. While the standard number of scans per pointings is three, thereby totaling a cadence of six scans per pair, some pointings were observed for just one or two scans. We note the sparsity of filter 3 candidates, specifically observing that all filter 3 events occur in pointings with fewer than three scans. This is likely due to the nature of filter 3, which selects events appearing in all ON scans–fewer scans increase the likelihood of detection in every scan.

| SRT K-band: 18.100 to 18.475 GHz | | | | | | | |
|---|---|---|---|---|---|---|---|
| Pair | ON-source pointing | RA (J2000) (h:m:s) | DEC (J2000) (d:m:s) | l (deg) | b (deg) | f2 Events | f3 Events |
| A00-D10 | A00 | 17:45:40.04 | −29:00:28.10 | 359.944 | −0.046 | 3 | 0 |
| | D10 | 17:45:34.28 | −29:02:31.77 | 359.904 | −0.046 | 3 | 0 |
| A00-D01 | A00 | 17:45:40.04 | −29:00:28.10 | 359.944 | −0.046 | 5 | 0 |
| | D01 | 17:45:45.79 | −28:58:24.41 | 359.985 | −0.046 | 4 | 0 |
| D03-D18 | D03 | 17:45:38.43 | −28:58:22.06 | 359.971 | −0.023 | 2 | 0 |
| | D18 | 17:45:47.56 | −28:59:06.81 | 359.978 | −0.058 | 1 | 0 |
| D02-D17 | D02[a] | 17:45:42.11 | −28:58:23.24 | 359.978 | −0.035 | 1 | 1 |
| | D17[b] | 17:45:49.32 | −28:59:49.20 | 359.971 | −0.069 | 4 | 0 |
| C01-D16 | C01[b] | 17:45:43.88 | −28:59:05.64 | 359.971 | −0.046 | 1 | 0 |
| | D16[b] | 17:45:51.08 | −29:00:31.60 | 359.964 | −0.081 | 2 | 0 |
| C02-C11 | C02[b] | 17:45:40.20 | −28:59:04.47 | 359.964 | −0.035 | 4 | 2 |
| | C11 | 17:45:47.40 | −29:00:30.44 | 359.958 | −0.069 | 13 | 0 |
| C03-C12 | C03 | 17:45:36.52 | −28:59:03.29 | 359.958 | −0.023 | 6 | 0 |
| | C12 | 17:45:45.64 | −28:59:48.04 | 359.964 | −0.058 | 6 | 0 |
| D05-B01 | D05[b] | 17:45:32.84 | −28:59:02.10 | 359.951 | −0.011 | 4 | 0 |
| | B01[b] | 17:45:42.00 | −28:59:46.87 | 359.958 | −0.046 | 4 | 0 |
| C04-B06 | C04[b] | 17:45:34.60 | −28:59:44.51 | 359.944 | −0.023 | 3 | 0 |
| | B06[b] | 17:45:43.72 | −29:00:29.27 | 359.951 | −0.058 | 1 | 0 |
| B02-C10 | B02[b] | 17:45:38.28 | −28:59:45.69 | 359.951 | −0.035 | 1 | 0 |
| | C10[b] | 17:45:45.48 | −29:01:11.67 | 359.944 | −0.069 | 4 | 0 |
| C05-B05 | C05 | 17:45:32.68 | −29:00:25.73 | 359.931 | −0.023 | 3 | 0 |
| | B05[b] | 17:45:41.80 | −29:01:10.50 | 359.938 | −0.058 | 3 | 0 |
| B03-C09 | B03[b] | 17:45:36.36 | −29:00:26.92 | 359.938 | −0.035 | 1 | 0 |
| | C09[b] | 17:45:43.57 | −29:01:52.91 | 359.931 | −0.069 | 1 | 0 |
| B04-D13 | B04[b] | 17:45:38.12 | −29:01:09.33 | 359.931 | −0.046 | 2 | 0 |
| | D13 | 17:45:45.33 | −29:02:35.31 | 359.924 | −0.081 | 2 | 0 |
| SRT K-band: 18.100 to 18.475 GHz | | | | | | | |
| C08-D15 | C08[a] | 17:45:39.88 | −29:01:51.73 | 359.924 | −0.058 | 5 | 5 |
| | D15[a] | 17:45:49.17 | −29:01:12.84 | 359.951 | −0.081 | 0 | 0 |
| C07-D14 | C07[b] | 17:45:36.20 | −29:01:50.55 | 359.917 | −0.046 | 1 | 0 |
| | D14[b] | 17:45:47.25 | −29:01:54.07 | 359.938 | −0.081 | 4 | 0 |
| C06-D12 | C06 | 17:45:34.44 | −29:01:08.14 | 359.924 | −0.035 | 5 | 0 |
| | D12 | 17:45:41.65 | −29:02:34.14 | 359.917 | −0.069 | 1 | 0 |
| D08-D11 | D08 | 17:45:30.76 | −29:01:06.96 | 359.917 | −0.023 | 4 | 0 |
| | D11 | 17:45:37.97 | −29:02:32.96 | 359.911 | −0.058 | 2 | 0 |
| D06-D09 | D06 | 17:45:30.92 | −28:59:43.32 | 359.938 | −0.011 | 10 | 0 |
| | D09 | 17:45:32.52 | −29:01:49.37 | 359.911 | −0.035 | 4 | 0 |
| D07-D04 | D07 | 17:45:29.00 | −29:00:24.54 | 359.924 | −0.011 | 2 | 0 |
| | D04 | 17:45:34.75 | −28:58:20.88 | 359.964 | −0.011 | 1 | 0 |

[a] Denote the pointings with one scan.

[b] Denote the pointings with two scans.

### 4.3. Constraints on the prevalence ETI transmitters

Although we did not find any convincing evidence of technosignatures at C-band and K-band across the observations towards 72 TESS targets and at the GC, our survey can still provide meaningful constraints on the prevalence of ETI transmitters operating at these frequencies in the Milky Way. The Drake equation [21,22] provides a simple metric to quantify the prevalence of transmitter rates. We can define the total number of operational extra-terrestrial transmitters that we can detect in the Milky Way as:

$$N = R_{IP} \, f_C \, L, \tag{3}$$

where $R_{IP}$ is the formation rate of technologically-advanced intelligent life that can build radio transmitters, $L$ is the length of time the transmitter is operational, and $f_C$ is the fraction of these transmitters that we are capable of detecting. The $R_{IP}$ is a composite term that includes various probability factors such as the rate of star formation, planet formation, and the likelihood of planets hosting life, and fraction of these life develop technology. Although we lack prior knowledge about $R_I P$ and $L$, based on the survey we conducted, we can constrain them together ($R_{IP} L \leq 1/f_C$) since we have placed limits on the fraction of stars that host detectable transmitters within our sensitivity limits. We carried out searches towards two groups of sources: the 72 stars from the TESS catalog that are in the solar neighborhood, and a region around the GC, which is farther away but likely contains a larger number of stars. Using the calculations provided by [15], we can estimate the rough number of stars in an $9' \times 9'$ region at C-band and a $5' \times 5'$ region at K-band in the GC bulge (see Fig. 3). We found $1 \times 10^6$ and $2.5 \times 10^5$ stars at C-band and K-band frequencies, respectively. We do not know the luminosity distribution of putative ETI transmitters in the Milky Way, but it is reasonable to assume that such a distribution would follow a power-law, with a relatively high number of low-power transmitters and a relatively small number of high-power transmitters, as is seen for terrestrial transmitters. This luminosity distribution can be expressed as:

$$N(EIRP) \propto EIRP^{-\alpha}, \tag{4}$$

where $\alpha$ is the slope. [3] compared a number of surveys conducted at lower frequencies around 1 GHz and reported a slope of 0.74. Since our study is one of the very few at 6 GHz and the first of its kind at the highest frequency of 18 GHz, we can derive the slope of this power-law for putative transmitters independently at our frequencies. Fig. 12 shows the inferred $f_C$ across a range of EIRP power levels for both groups of stars combined from both bands. As evident from Fig. 12, our study finds the slope of the luminosity function of putative transmitters is ≈0.99, which is steeper compared to the slope of luminosity distribution constrained by [3]. Hence, based on this survey, we can conclude that fewer than 1 in $10^3$ stars host narrowband transmitters with power levels reaching that of a Kardashev I civilization.

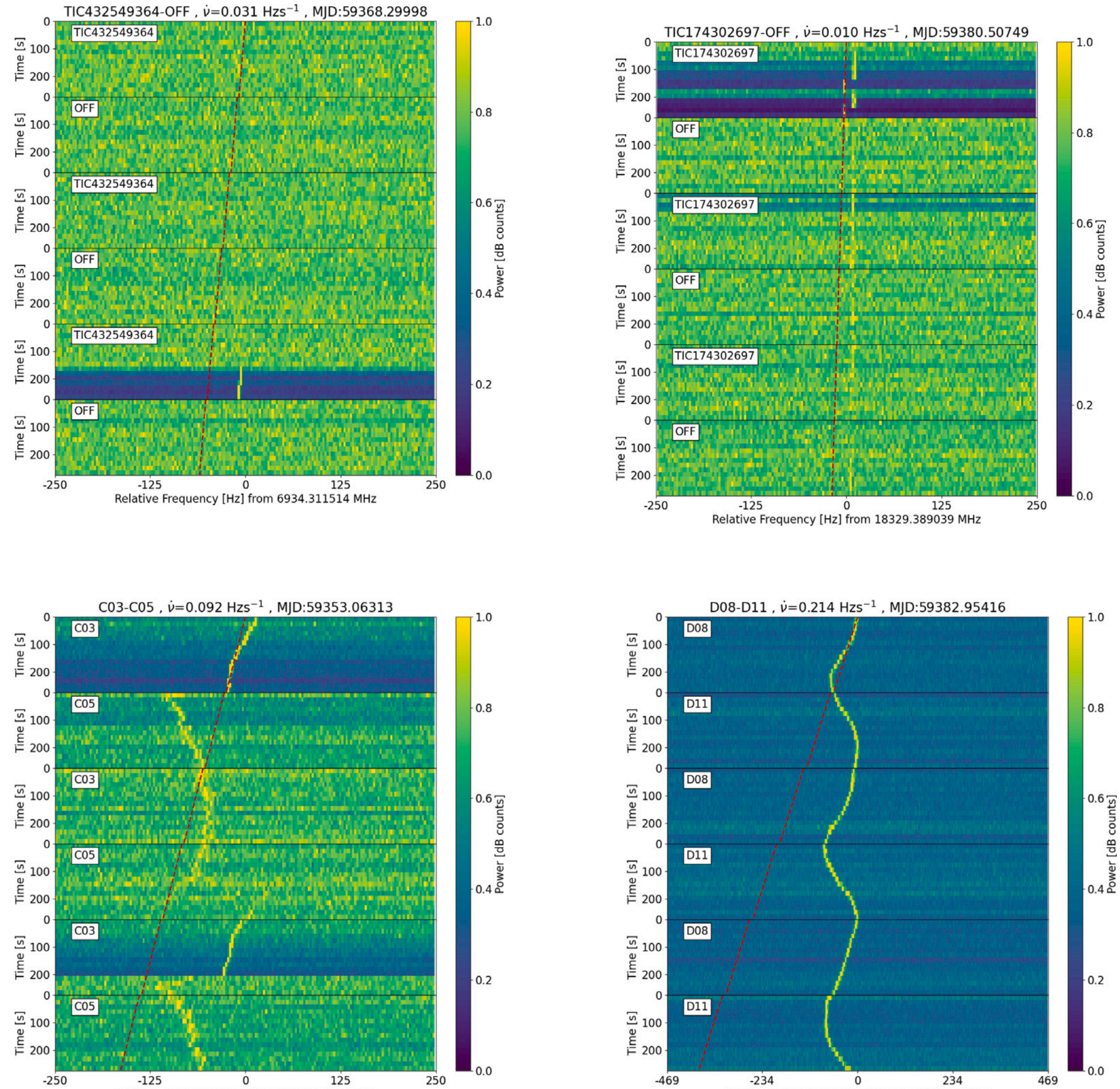

**Fig. 11.** Examples of filter 2 narrowband drifting signals found from the TESS and GC SRT survey conducted at C- and K-band, respectively. For each plot, dynamic spectra are shown for six observing ON-OFF pairs with the top panel showing the first scan in the pair. The top of the plot shows the field names for both pointings of the pair. The center frequency of each event and measured drift rate are also listed at the bottom and top of each plot, respectively. The overlaid red dashed line shows the detected drift rate obtained from the first ON observation in which the signal first appears with a slight offset in frequency for visualization. The Top left plot Shows a TESS C-band event, and the Top right one shows a TESS K-band event. Down left one shows a GC C-band event, and the Down right one shows a K-band GC event. We reject these candidates as RFI on the basis of their non-repeating nature and/or their presence in the OFF scans.

It should be noted that our analysis relies on two key assumptions. First, our pipeline employs an 18-s averaging time via Fast Fourier Transform to produce high-spectral resolution data products. This assumes that the underlying signal persists for a significant fraction of this 18-s window. If the signal is highly intermittent or pulsates on timescales much shorter than 18 s (i.e., with a very low duty cycle), our averaging will lead to a sensitivity loss approximated by $EIRP_{\rm min}/\sqrt{D}$, where $D$ is the duty cycle. Second, our signal detection criteria require that a signal be continuously present throughout the entire 30 min observation period (i.e., for filter 3 hits). Consequently, our selection criteria reduce sensitivity to transient, intermittent, or short-duration periodic signals that do not persist for the full 30 min. As a result, certain classes of ETI transmissions—such as brief bursts or low duty cycle signals—might be partially averaged out and remain undetected. Future work incorporating shorter integration times or alternative signal processing techniques could help address this limitation.

Additionally, we can determine an upper limit on the approximate number of narrowband ETI transmitters likely to exist in the entire Milky Way. Let the prevalence rate of ETI transmitters in the Milky Way galaxy be denoted as $\lambda$. Given the extremely low expected occurrence rate, we assume that this follows a Poisson process [5,6,23], which is described by:

$$P(k) = \frac{\lambda^k e^{-\lambda}}{k!}, \tag{5}$$

where $k$ is the number of detected signals. In our survey, since no transmitters were detected ($k = 0$), we can estimate an upper limit on $\lambda$ with a 95% confidence interval as:

$$\lambda = -\ln(0.05) \approx 3. \tag{6}$$

In our survey of TESS targets covering 72 stars, and the GC region covering approximately $10^5$ stars across both frequency bands, we can estimate the probability of detecting a transmitter ($p_t$), assuming such transmitters exist, using the formula:

$$p_t = \frac{\lambda}{N_{\rm targets}}. \tag{7}$$

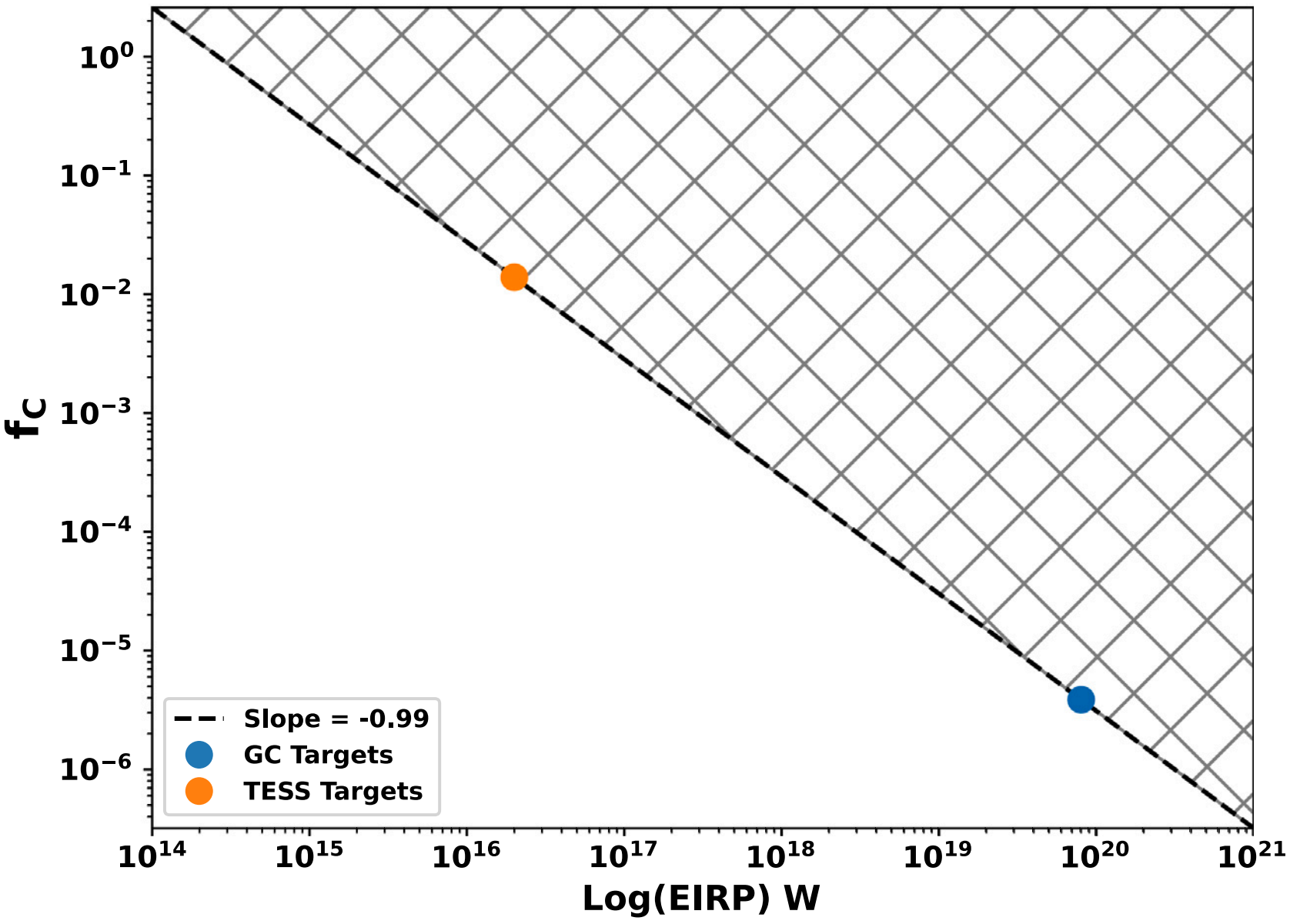

**Fig. 12.** Transmitter-Rate constraints achieved from our survey of nearby TESS sources, combined with the expected number of stars at the Galactic Center. The $X$-axis represents EIRP in watts, while the $Y$-axis denotes the transmitter rate. For simplicity, we have combined constraints from both of our bands for each of these sets of targets. The shaded region indicates the parameter space that our survey could conclusively reject as hosting transmitters. As this survey is the first ever conducted at these frequencies, it establishes some of the initial constraints on the existence of ETI transmitters and their rates at 18 GHz.

For the nearby TESS stars, the probability of detecting a transmitter was approximately 3.9%, while for the GC region, the probability was around 0.003%. Extrapolating these results, we estimate that, assuming the Milky Way contains $10^{11}$ stars, the upper limit on the number of stars that could host detectable narrowband transmitters with an EIRP $\geq 10^{19}$ W is $N(>EIRP) \leq 3 \times 10^6$ at the observed survey frequencies. Advanced machine learning could significantly enhance the detection of weaker or high-drift signals, which may be missed with traditional methods. Future work will explore AI-driven solutions to expand the search parameter space and improve detection capabilities, especially for weak signals under the EIRP limits, such as those corresponding to a Kardashev Type I civilization.

### 4.4. *Highest frequency survey for ETI transmitters*

We detected more than 156,156 raw hits in the C-band and approximately 16,015 raw hits in the K-band from both sets of targets. To facilitate comparison, we normalized these numbers by the total observing time and observed bandwidth. We found that at C-band, the normalized raw hit rate was 1.5 hits/MHz/h, while at K-band, it was approximately 1.8 hits/MHz/h. This suggests that although the absolute number of raw hits at C-band was an order of magnitude higher, both bands exhibit a similar number of normalized hits. However, there is a significant disparity between the number of raw hits detected at lower frequencies compared to higher frequencies. For instance, Choza et al. [6] conducted observations of nearby galaxies across 1–10,GHz. Normalizing their raw hit counts using the same method, we find that their normalized hit rate at L-band (1.4 GHz) was approximately 54.6 hits/MHz/h, whereas at C-band, it was 7 hits/MHz/h—an order of magnitude difference. Similarly, using observations reported by Traas et al. [5], the normalized hit rate was approximately 24 hits/MHz/h at L-band compared to only 1.0 hits/MHz/h at C-band, which is consistent with our findings at C-band. These results indicate that conducting technosignature searches at higher frequencies (>4 GHz) significantly reduces the false positive rate by an order of magnitude.

Radio technosignature searches are multidimensional, with several unknown parameters. The direction in the sky where a putative transmitter might be located, and the frequency at which such a transmitter might be operating, are two of the most crucial, with the signal type being the third. As discussed in Section 1, narrowband signals are an obvious choice, but efforts are underway to cover other classes of signal types as well [24,25]. To account for the unknown parameters, searches must cover a large number of stars across the entire electromagnetic spectrum, especially the radio portion accessible through the atmosphere, which spans from 1 to 100 GHz. Fig. 13 shows various surveys conducted over the years and their respective frequency ranges, highlighting the number of stars covered at their respective frequencies. It is clearly evident, that by comparing our survey to all the previous surveys, our work is one of the highest-frequency technosignature surveys ever conducted, covering the largest number of stars at these frequencies.

## 5. Conclusions

We report one of the first radio technosignature searches at 6.5 GHz and 18 GHz, with the latter constituting the highest-frequency survey to date (1–100 GHz) targeting over $10^5$ stars searching for the narrowband signals (<1 Hz). We also report the first radio technosignature searches from the Sardinia Radio Telescope, which is equipped to carry out observations up to 115 GHz. We conducted 29 h of observations across 5.7–7.7 GHz in 19 different pointings towards the GC and 21 h of observations towards 42 TESS TOIs across 6.57–6.95 GHz. Additionally, at higher frequencies, we conducted 15 h of observations towards 30 TOIs across 18.1–18.5 GHz and 8 h of observations towards 37 pointings in the GC region across 18.08–18.46 GHz. We performed detailed searches for narrowband drifting signals but did not find any convincing evidence of technosignatures in our 73 h of combined observations across both bands and sets of targets. Through our survey, we were able to constrain the existence of a narrowband radio transmitter characterized by an high duty cycle ($\sim$1) operating at the GC with an EIRP of $1.4 \times 10^{20}$ W at C-band and at K-band with an EIRP of $8 \times 10^{19}$ W. Similarly, in the solar neighborhood ($\lesssim$272 pc), we were able to constrain the existence of a narrowband transmitter with an EIRP of $3.7 \times 10^{16}$ W at C-band and at K-band with an EIRP of $2 \times 10^{16}$ W. Overall, we establish one of the first constraints on the number of narrowband transmitters at these frequencies in the Milky Way, finding $N(>EIRP) \leq 3 \times 10^6$ for EIRP $\geq 10^{19}$ W. Future surveys at even higher frequencies with larger instantaneous bandwidth aided by AI-enabled tools, covering more stars, will be able to further constrain these upper limits, and the Sardinia Radio Telescope will be at the forefront of pushing these boundaries.

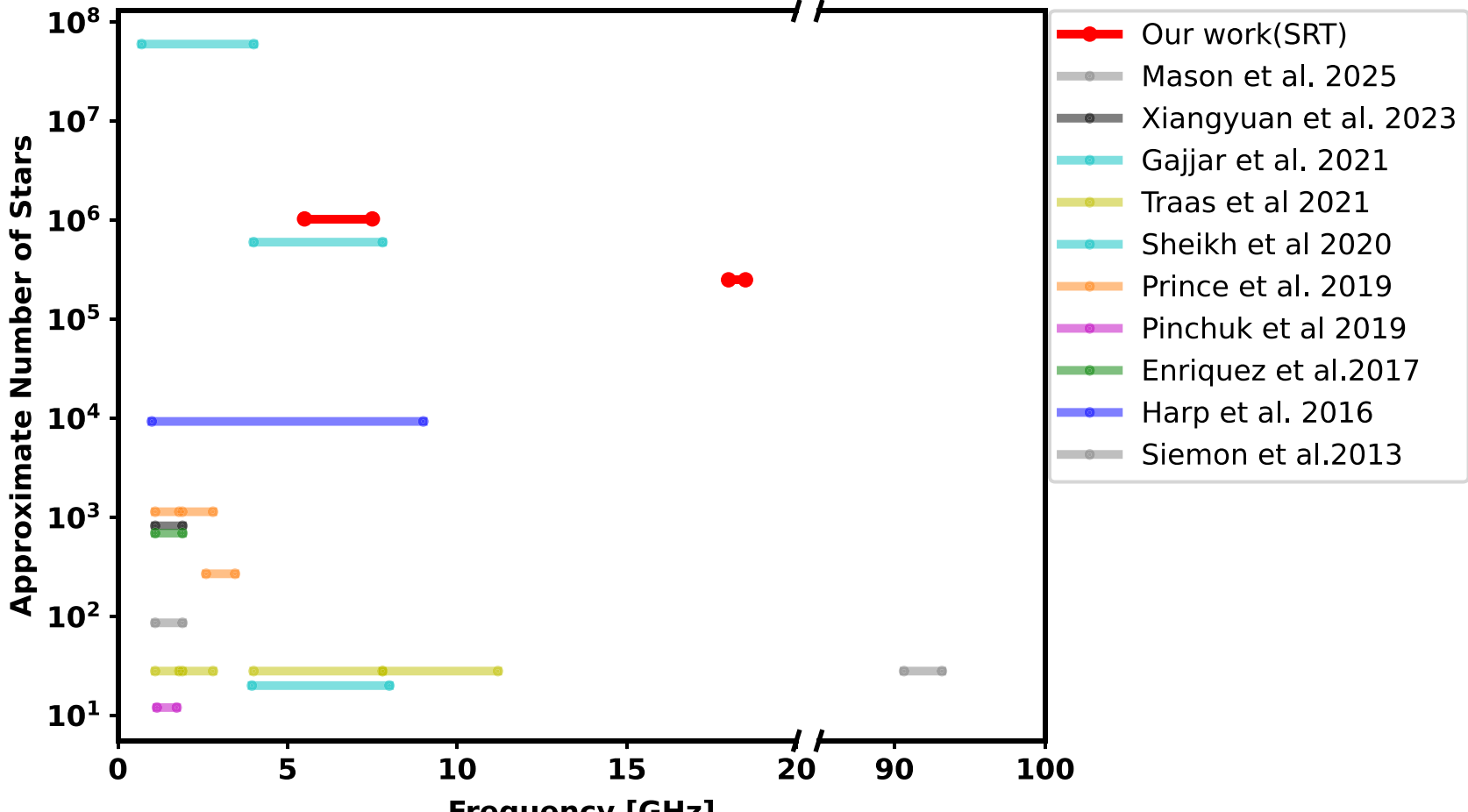

**Fig. 13.** Comparison of survey frequency with the number of stars surveyed for a collection of previous radio technosignature surveys. To date, large fraction of surveys have been conducted at frequencies below 12 GHz. Thus, we present one of the very few constraints on a putative ETI transmitter operating at some of the highest radio frequencies.

## CRediT authorship contribution statement

**Lorenzo Manunza:** Data curation. **Alice Vendrame:** Data curation. **Luca Pizzuto:** Data curation. **Monica Mulas:** Data curation. **Karen I. Perez:** Data curation. **Vishal Gajjar:** Data curation. **Andrea Melis:** Supervision, Investigation, Data curation. **Maura Pilia:** Supervision, Investigation, Data curation. **Delphine Perrodin:** Data curation. **Giambattista Aresu:** Data curation. **Marta Burgay:** Data curation. **Alessandro Cabras:** Data curation. **Giuseppe Carboni:** Data curation. **Silvia Casu:** Data curation. **Tiziana Coiana:** Data curation. **Alessandro Corongiu:** Data curation. **Steve Croft:** Data curation. **Elise Egron:** Data curation. **Owen A. Johnson:** Data curation. **Adelaide Ladu:** Data curation. **Matt Lebofsky:** Data curation. **Francesca Loi:** Data curation. **David MacMahon:** Data curation. **Carlo Migoni:** Data curation. **Emilio Molinari:** Data curation. **Matteo Murgia:** Data curation. **Alberto Pellizzoni:** Data curation. **Tonino Pisanu:** Data curation. **Antonio Poddighe:** Data curation. **Andrea Possenti:** Data curation. **Erika Rea:** Data curation. **Andrew Siemion:** Data curation. **Paolo Soletta:** Data curation. **Matteo Trudu:** Data curation. **Valentina Vacca:** Data curation.

## Declaration of competing interest

The authors declare that they have no known competing financial interests or personal relationships that could have appeared to influence the work reported in this paper.

## Acknowledgments

We are thankful to the director of the INAF Astronomical of Cagliari for the Discretionary Directorial Time request with the Sardinia Radio Telescope. We also thank the INAF Scientific Direction that kindly financed our INAF activities up to 2025. The Breakthrough Prize Foundation funds the Breakthrough Initiatives which manages Breakthrough Listen.